\documentclass[twocolumn,english,british,prl, showpacs]{revtex4}
\usepackage[latin9]{inputenc}
\usepackage{amsmath}
\usepackage{graphicx}

\makeatletter
\@ifundefined{textcolor}{}
{%
 \definecolor{BLACK}{gray}{0}
 \definecolor{WHITE}{gray}{1}
 \definecolor{RED}{rgb}{1,0,0}
 \definecolor{GREEN}{rgb}{0,1,0}
 \definecolor{BLUE}{rgb}{0,0,1}
 \definecolor{CYAN}{cmyk}{1,0,0,0}
 \definecolor{MAGENTA}{cmyk}{0,1,0,0}
 \definecolor{YELLOW}{cmyk}{0,0,1,0}
}

\makeatother

\usepackage{babel}
\begin{document}

\title{Dipolar broadening of nuclear spin resonance under dynamical pumping}

\author{O. Tsyplyatyev}

\affiliation{Department of Physics and Astronomy, University of Sheffield, Sheffield
S3 7RH, UK}

\author{D. M. Whittaker}

\affiliation{Department of Physics and Astronomy, University of Sheffield, Sheffield S3 7RH, UK}
\begin{abstract}
We study  the polarisation dependence of the homogeneously broadened
nuclear spin resonance in a crystal. We employ a combinatorial method to restrict
the nuclear states  to a fixed polarisation
and show that the centre of the resonance is shifted linearly with
the nuclear polarisation by up to the zero polarisation line width.
The width shrinks from its maximum value at zero polarisation to zero
at full polarisation. This  suggests to use the line  shape  as a  direct measure of nuclear polarisation reached under dynamical pumping.  In the limit of single quantum of excitation above
the fully ferromagnetic state, we provide  an explicit solution to the problem of nuclear spin dynamics which  links a bound on the fastest decay rate to the observable width of the resonance line.

\end{abstract}

\pacs{76.20.+q, 76.60.-k, 75.10.Jm}

\maketitle

\section{Introduction}

Recent interest in studying nuclear spins was fuelled
\cite{Bluhm,Reilly,Tarucha,KhaetskiiLossGlazman,Tsyplyatyev,Pedrocchi,Stepanenko,Petta}
by the dephasing effect in GaAs quantum dots which is induced by fluctuations of the nuclear (Overhauser) field  and blocks prospective application of electron spins  for  spintronics
\cite{DasSarma} and quantum computing \cite{LossDivicenzo}. Among
different approaches to this problem, there is a new tool that combines  radio
frequency excitation and  dynamical nuclear pumping techniques to
control the Overhauser field and simultaneously to detect its state \cite{Gammon}. 
In particular,  this method allows to  access directly some of  the local properties of the nuclear bath.

At infinite temperature, a connection between  the line width of the nuclear magnetic resonance and  the dipole interaction between nuclei  in a
crystal was established in \cite{vanVleck}. Later, it was found that
the line shape changes at  low spin temperatures \cite{Abragam}. 
A detailed understanding of the effect of a non-equilibrium polarisation, as occurs in dynamical pumping experiments,  may give a  way to measure directly the degree of polarisation reached under dynamical pumping and provide  further insights in the effect of dipole interaction  which plays an important role in the intrinsic
dynamics of the nuclear bath in unstrained structures.

In present paper we address this problem by analysing the long wavelength spectrum of polarised
nuclei using a cumulant expansion of the full line shape. We employ
a combinatorial method to restrict the nuclear states to a fixed polarisation
$p=S_{z}/\left(IN\right)$, where  $S_{z}$ is the number of excitation
quanta pumped to the nuclear bath, and find that the centre of the
resonance shifts linearly with the polarisation, 
\begin{equation}
\left\langle \nu\right\rangle =\mu B+3IFp\label{eq:nu_I_1}
\end{equation}
where $B$ is the external magnetic field, $\mu$ is the nuclear magneton, $I$ is spin of a nucleus, $N$
is the total number of nuclei,
and $F$ is a maximum field induced by fully polarised nuclear bath.
The line width at a finite polarisation is reduced, relative to an unpolarised bath, as 
\begin{equation}
\left\langle \nu^{2}\right\rangle =3I\left(I+1\right)F_{2}w\left(p\right),\label{eq:nu_I_2}
\end{equation}
where $F_{2}$ is a fluctuating field induced by completely unpolarised
nuclear bath and $w\left(p\right)$ is a bell shaped function, $w\left(0\right)=1$
and $w\left(\pm1\right)=0$; $w\left(p\right)=1-p^{2}$ for $I=1/2$
and the line width shrinks slightly faster for $I>1/2$. The third
cumulant is zero when $p=0$ but becomes comparable to the line width
at a moderate polarisation signalling the build-up of a line shape asymmetry.

One way to suppress the dephasing effect of the Overhauser field is to polarise the nuclei above 99\% \cite{KhaetskiiLossGlazman}. So far the polarisation reached in dynamical nuclear pumping experiments is estimated indirectly via the hyperfine shift of the electron spin and, as reported, is still below the required threshold \cite{Reilly}. 
The result of this paper suggests to use the shape of the nuclear magnetic resonance as a  direct indicator of the degree of polarisation achieved in  quantum dots. 

The vanishing of the broadening at large polarisations is caused by
predominance of the exchange part of the dipole interaction. In the
limit of single quantum of excitation above the ferromagnetic state
the eigenmodes are plane waves with a sinusoidal dispersion, which allows
to solve the problem of nuclear spin dynamics explicitly. We show that 
the observed line width of the nuclear resonance in a single crystal of GaAs \cite{barrett}
gives a bound on the shortest polarisation decay times for lateral quantum dots as $T_{1}>1.25$
s which is only an order of magnitude shorter than the $T_{1}=10-100$ s measured in these dots \cite{Reilly,Tartakovskii} at $p\ll1$. Note that, in this bound, we neglect quadrupole and hyperfine interactions of  nuclear spins. Thus it is valid for unstrained structures without conduction or localised electrons.

The paper is organised as follows. In Section II we approximate the model of dipole interaction between nuclear spins for a large external magnetic field. In Section III we analyse the shape the absorption line to obtain its polarisation dependence. Section IV contains the analysis of single quantum of excitation above the fully ferromagnetic state where we establish a bound on the shortest poralisation decay time. In the appendices we give some details of the calculations in Section III.

\section{Dipole interaction}
The Hamiltonian that describes $N$ nuclear spins located at the sites
of a regular lattice, their dipole interaction with the strength $g$,
and an external magnetic field $B$ is given by 
\begin{equation}
H=\mu B\sum_{j}I_{j}^{z}+\sum_{i<j}\frac{g}{r_{ij}^{3}}\left(3\left(\mathbf{e}_{ij}\cdot\mathbf{I}_{i}\right)\left(\mathbf{e}_{ij}\cdot\mathbf{I}_{j}\right)-\mathbf{I}_{i}\cdot\mathbf{I}_{j}\right),
\end{equation}
where $\mu$ is the nuclear magneton, $I_{j}^{z},I_{j}^{\pm}=I_{j}^{x}\pm iI_{j}^{y}$
are the spin-$I$ operators, index $j$ labels the sites on the lattice,
$r_{ij}$ is the distance between two sites, and $\mathbf{e}_{ij}=\mathbf{r}_{ij}/r_{ij}$
is the unit vector that connects sites $i$ and $j$. We consider
a $d$-dimensional crystal with $N=L^{d}$ nuclear spins and assume
periodic boundary conditions, $\mathbf{I}_{j+L}=\mathbf{I}_{j}$,
to exclude edge states from the analysis.

In a strong magnetic field the Hilbert space is partitioned by a large
Zeeman energy into a set of subspaces labelled by the total z-projection
of all spins, $S_{z}$ is an eigenvalue of $\sum_{j}I_{j}^{z}$. In
each subspace the  many spin states that have the same Zeeman energy
are split by a part of the dipole interaction that conserves $S_{z}$, 

\begin{equation}
H_{0}=\mu B\sum_{j}S_{j}^{z}+\sum_{i<j}\frac{g\left(3\gamma_{ij}^{2}-1\right)}{r_{ij}^{3}}\left(I_{j}^{z}I_{i}^{z}-\frac{I_{j}^{+}I_{i}^{-}}{2}\right).\label{eq:H0}
\end{equation}
The matrix elements of the $S_{z}$ non-conserving remainder, $H_{1}=\sum_{ij}3g\Big[\left(\alpha_{ij}+i\beta_{ij}\right)^{2}I_{j}^{+}I_{i}^{+}+4\left(\alpha_{ij}+i\beta_{ij}\right)\gamma_{ij}I_{i}^{+}I_{j}^{z}+\left(\alpha_{ij}-i\beta_{ij}\right)^{2}I_{j}^{-}I_{i}^{-}+4\left(\alpha_{ij}-i\beta_{ij}\right)\gamma_{ij}I_{i}^{-}I_{j}^{z}\Big]/\left(8r_{ij}^{3}\right),$
connect subspaces with different Zeeman energies and can be treated
perturbatively when the magnetic field is large.
Here $\left(\alpha_{ij},\beta_{ij},\gamma_{ij}\right)$ are coordinates
of the vector $\mathbf{e}_{ij}$ in a reference frame with z-axis
parallel to the external $B$-field.

\section {Absorption line shape}
The absorption of electromagnetic radiation in a strong magnetic field
is dominated by a flip of a single nuclear spin at an energy of $\mu B$, neglecting the dipole interaction. The dipole interaction of a single spin with other nuclear spins on
the lattice makes the width of the resonance finite. Following \cite{vanVleck}
we study moments of this line shape given by the transition energies
$E_{f}-E_{f'}$ between pairs of the eigenstates $\left|f\right\rangle $
and $\left|f'\right\rangle $, 
\begin{equation}
\left\langle m^{k}\right\rangle =\frac{\sum_{ff'}\left(E_{f'}-E_{f}\right)^{k}\left|\left\langle f'|I^{+}|f\right\rangle \right|^{2}}{\sum_{ff'}\left|\left\langle f'|I^{+}|f\right\rangle \right|^{2}},\label{eq:mk}
\end{equation}
where $I^{+}=\sum_{j}I_{j}^{+}$ is the operator that corresponds
to a long wave-length photon and $k$ is the order of the moment.
To account for a finite polarisation, the initial states $f$ are
restricted to a subspace of fixed $S_{z}$.
In our study of the dipolar broadening we will use only the truncated
part $E_{f}\left|f\right\rangle =H_{0}\left|f\right\rangle $ of the
dipole Hamiltonian $H$. A perturbative treatment of $H_{1}$ produces
extra peaks at $2\mu B$ and $3\mu B$ with small amplitudes $\sim g/B$,
which are neglected.

The sums in Eq. (\ref{eq:mk}) can be transformed to traces by means
of general quantum mechanical identities. In the denominator, the
sum over the eigenstates $\left|f'\right\rangle $ gives a unit matrix,
$\sum_{f'}\left|f'\left\rangle \right\langle f'\right|=\hat{1}$,
in the subspace of $S^{z}+1$ as the operator $I^{+}$ changes $S_{z}$
by 1 only. Then, the remaining sum over
a complete set of the eigenstates is $\sum_{f}\left\langle f|I^{-}I^{+}|f\right\rangle =\textrm{Tr}\left(I^{-}I^{+}\right)$
where the trace is restricted to a subspace of fixed $S^{z}$. Representing
the eigenenergies as $E_{f}=\left\langle f|H_{0}|f\right\rangle $
the numerator can be transformed in a similar way. Explicit expressions
for the first three moments are
\begin{eqnarray}
\left\langle m\right\rangle  & = & \frac{-\textrm{Tr}\left(\left[H_{0},I^{-}\right]I^{+}\right)}{\textrm{Tr}\left(I^{-}I^{+}\right)},\nonumber \\
\left\langle m^{2}\right\rangle  & = & \frac{-\textrm{Tr}\left(\left[H_{0},I^{-}\right]\left[H_{0},I^{+}\right]\right)}{\textrm{Tr}\left(I^{-}I^{+}\right)},\label{eq:m123}\\
\left\langle m^{3}\right\rangle  & = & \frac{\textrm{Tr}\left(\left[H_{0},\left[H_{0},I^{-}\right]\right]\left[H_{0},I^{+}\right]\right)}{\textrm{Tr}\left(I^{-}I^{+}\right)},\nonumber 
\end{eqnarray}
where $\left[A,B\right]=AB-BA$ is the operator commutator. Note, for calculating the absorption spectrum, that $I^{+}$has to appear on the right of $I^{-}$ in the traces.

\subsection{$I=1/2$ case}
As a trace is invariant with respect to a change of basis we will
calculate Eq. (\ref{eq:m123}) in the basis of non-interacting spins
that are quantised individually along the external magnetic field.
First, let us consider the nuclei with spin $I=1/2$. In the basis
of non interacting states,
the denominator is a single sum of traces of individual spin operators
$\sum_{ij}\textrm{Tr}\left(I_{i}^{-}I_{j}^{+}\right)=\sum_{i}\textrm{Tr}\left(I_{i}^{-}I_{i}^{+}\right)$.
The full set of the basis states in a subspace of fixed $S^{z}$ can
be divided in two groups: one has the $i^{th}$ spin in the state
$I_{i}^{z}=1/2$ and the other in the state $I_{i}^{z}=-1/2$. For
each state form the first group the expectation value of $I_{i}^{-}I_{i}^{+}$
is $0$ and for the second group it is $1$. Then, the trace of every
single spin operator is independent of $i$, $\textrm{Tr}\left(I_{i}^{-}I_{i}^{+}\right)=0\cdot C_{N-1}^{n-1}+1\cdot C_{N-1}^{n}$, where the binomial factors $C_{N-1}^{n-1}$ and $C_{N-1}^{n}$ give the numbers of states in each group. Here $n$ is
the number of 1/2-spins pointing up with respect to the fully polarised
state, i. e. $S^{z}=-N/2+n$. The sum over all $N$ single spin traces
gives the denominator as
\begin{equation}
\textrm{Tr}\left(I^{-}I^{+}\right)=\frac{N!}{n!\left(N-n-1\right)!}.\label{eq:denominator_12}
\end{equation}

The numerator of the first moment $k=1$ requires only one first order
commutator, $\left[H_{0},I^{-}\right]=-\mu B\sum_{j=1}^{N}I_{j}^{-}-\sum_{i<j}\frac{g}{r_{ij}^{3}}3\left(3\gamma_{ij}^{2}-1\right)I_{j}^{z}I_{i}^{-}.$
The trace of a two spin operator, that is involved in Eq. (\ref{eq:m123}),
is evaluated analogously to the denominator. We calculate averages
over all possible two spin states accounting for the remaining $N-2$
spin states by  binomial factors,
\begin{equation}
\textrm{Tr}\left(I_{j}^{z}I_{i}^{-}I_{i}^{+}\right)=\frac{\left(1-\delta_{ij}\right)}{2}\frac{\left(N-2\right)!\left(N-2n-1\right)}{n!\left(N-n-1\right)!}.\label{eq:tr_2_12}
\end{equation}
Substituting the commutator $\left[H_{0},I^{-}\right]$ into Eq. (\ref{eq:m123})
and using the last expression we obtain the line shift, $\left\langle \nu\right\rangle =\left\langle m\right\rangle $,
that corresponds to the first moment 
\begin{equation}
\left\langle \nu\right\rangle =\mu B-F\frac{3}{2}\frac{N-2n-1}{N-1},
\end{equation}
where $F=g\sum_{j}\left(3\gamma_{ij}^{2}-1\right)/r_{ij}^{3}$ is
independent of the position on the lattice $i$ due to the periodic
boundary conditions. Expanding the above expression in a $1/N$-series
and identifying the polarisation as $p=-1+2n/N$ we obtain Eq. (\ref{eq:nu_I_1})
for the case  $I=1/2$.

The numerator of the second moment $k=2$ requires
an extra first order commutator $\left[H_0,I^{+}\right]=B\sum_{j}I_{j}^{+}+\sum_{i<j}\frac{g}{r_{ij}^{3}}3\left(3z_{ij}^{2}-1\right)I_{j}^{+}I_{i}^{z}$.
The trace of a three spin operator, which is needed for $\left\langle m^{2}\right\rangle $,
is obtained in a similar way,
\begin{multline}
\textrm{Tr}\left(I_{j}^{z}I_{i}^{-}I_{i}^{+}I_{i'}^{z}\right)=\frac{\left(1-\delta_{ij}\right)\left(1-\delta_{ii'}\right)\left(N-3\right)!}{4n!\left(N-n-1\right)!}\\
\left[(2n-N+1)(2n-N+2)-\delta_{ii'}2n(2n-2N+3)\right].\label{eq:tr_3_12}
\end{multline}
Substituting the product of the two commutators $\left[H,I^{-}\right]\left[H,I^{+}\right]$
into Eq. (\ref{eq:m123}), and using the above expression, we obtain
the second cumulant $\left\langle \nu^{2}\right\rangle =\left\langle m^{2}\right\rangle -\left\langle m\right\rangle ^{2}$
that corresponds to the line width, 
\begin{eqnarray}
\left\langle \nu^{2}\right\rangle  & = & F_{2}\frac{3n(2N-2n-3)}{2\left(N-1\right)\left(N-2\right)},
\end{eqnarray}
where $F_{2}=g^{2}\sum_{j}\left(3\gamma_{ij}^{2}-1\right)^{2}/r_{0j}^{6}$.
In the limit $N\gg1$ this expression leads to Eq. (\ref{eq:nu_I_2})
for the case $I=1/2$ and $w\left(p\right)=1-p^{2}$. 

The third moment contains an admixture of the first and the second
moments when the polarisation is finite, $p\neq0$. Hence, we evaluate the
third irreducible moment as $\left\langle \nu^{3}\right\rangle =\left\langle m^{3}\right\rangle -3\left\langle \nu^{2}\right\rangle \left\langle \nu\right\rangle -\left\langle \nu\right\rangle ^{3}$
using the same procedure as for the moments above. In the limit $N\gg1$,
see Appendix A, we obtain
\begin{equation}
\left\langle \nu^{3}\right\rangle =\frac{9}{8}\left(F_{3c}-F_{2}F-5F_{3}\right)p\left(1-p^{2}\right)\label{eq:nu_3_12}
\end{equation}
where $F_{3}=g^{3}\sum_{j}\left(3\gamma_{ij}^{2}-1\right)^{3}/r_{ij}^{9}$
and $F_{3c}=g^{3}\sum_{ij}\left(3\gamma_{lj}^{2}-1\right)\left(3\gamma_{ij}^{2}-1\right)\left(3\gamma_{li}^{2}-1\right)/\left(r_{lj}^{3}r_{ij}^{3}r_{il}^{3}\right)$.
This result indicates that the line shape is asymmetric when the nuclear
spins are polarised. The asymmetry becomes pronounced when the skewness
$\left\langle \nu^{3}\right\rangle /\left\langle \nu^{2}\right\rangle ^{3/2}=\sqrt{3/F_{2}^{3}}\left(F_{3c}-F_{2}F-5F_{3}\right)p/\sqrt{1-p^{2}}$
is of the order of one. For example, in a cubic crystal with external
magnetic field $\mathbf{B}\parallel\left[001\right]$ \cite{skewness_cubic}
this point is reached at a moderate polarisation of $p=0.32$.

\subsection{$I>1/2$ case}

For nuclei with $I>1/2$, the extension of the combinatorial calculation 
presents an extra difficulty due to the larger number of single spin states.
We start from the denominator of Eq. (\ref{eq:m123}) which involves
a trace over only one single spin operator, $\textrm{\textrm{Tr}}\left(I_{j}^{-}I_{j}^{+}\right)$.
Summing over $2I+1$ single spin states, instead of just $I_{i}^{z}=\pm1/2$,
modifies Eq. (\ref{eq:denominator_12}) to
\begin{equation}
\textrm{Tr}\left(I^{-}I^{+}\right)=N\sum_{k=0}^{2I}{}_{I}A_{n-k}^{N-1}\left(k+1\right)\left(2I-k\right),\label{eq:Z_A}
\end{equation}
where $_{I}A_{n}^{N}=\sum_{k=0}^{n/\left(2I+1\right)}\left(-1\right)^{k}C_{k}^{N}C_{n-\left(2I+1\right)k}^{N-1+n-\left(2I+1\right)k}$ 
 is the number of linearly independent eigenstates of the operator  $\sum_{j}I_{j}^{z}$ with $S_z=-NI+n$, constructed out of
$N$ spins $I$, see Appendix B. This provides a generalisation of the binomial factor
for $I>1/2$; it can be check that $_{\frac{1}{2}}A_{n}^{N}=C_{N}^{n}$.

The traces in Eqs. (\ref{eq:tr_2_12}, \ref{eq:tr_3_12}) of the two
and the tree single spin operators which are needed for the first
two moments are modified in the same way, using $_{I}A_{n}^{N}$ instead
of the binomial factor, 
\begin{multline}
\textrm{Tr}\left(I_{j}^{z}I_{i}^{-}I_{i}^{+}\right)=\left(1-\delta_{ij}\right)\sum_{k_{1},k_{2}=0}^{2I}{}_{I}A_{n-k_{1}-k_{2}}^{N-2}\\
\times\left(-I+k_{1}\right)\left(k_{2}+1\right)\left(2I-k_{2}\right)\label{eq:2spin_A}
\end{multline}
\begin{widetext}
\begin{multline}
\textrm{Tr}\left(I_{j}^{z}I_{i}^{-}I_{i}^{+}I_{i'}^{z}\right)=\delta_{ji'}\left(1-\delta_{ij}\right)\left(1-\delta_{ii'}\right)\sum_{k_{1},k_{2}=0}^{2I}{}_{I}A_{n-k_{1}-k_{2}}^{N-2}\left(k_{1}+1\right)\left(2I-k_{1}\right)\left(-I+k_{2}\right)^{2}\\
+\left(1-\delta_{ji'}\right)\left(1-\delta_{ij}\right)\left(1-\delta_{ii'}\right)\sum_{k_{1},k_{2},k_{3}=0}^{2I}{}_{I}A_{n-k_{1}-k_{2}-k_{3}}^{N-3}\left(k_{1}+1\right)\left(2I-k_{1}\right)\left(-I+k_{2}\right)\left(-I+k_{3}\right)\left(1-\delta_{ii'}\right)\label{eq:3spins_A}
\end{multline}
\end{widetext}
We substitute Eqs. (\ref{eq:Z_A},\ref{eq:2spin_A},\ref{eq:3spins_A})
into Eq. \foreignlanguage{english}{(\ref{eq:m123})} and evaluate
the cumulants numerically. The first cumulant
coincides with Eq. (\ref{eq:nu_I_1}) in the limit of a large crystal
$N\gg1$, while the second cumulant gives the Eq. (\ref{eq:nu_I_2}) with
$w\left(p\right)$ evaluated in Appendix C.

In a crystal with two species of nuclear spins the dipolar broadening
of the like spins has an extra contribution from the unlike spins.
We assume that the corresponding magnetons, $\mu_{a}$ and $\mu_{b}$,
are significantly different, $\left|\mu_{a}-\mu_{b}\right|B\gg g_{a,},g_{b},g_{ab}$
where $g_{a,},g_{b},g_{ab}$ are the dipolar coupling constants between the like and the unlike nuclei. Then the off-resonant interaction between the unlike nuclei makes the exchange term of Eq. (\ref{eq:H0}) a perturbation, which we will neglect. For the unlike spins we repeat the same calculation as for the like spins keeping only the Ising term of the dipolar interaction between the
two sublatices. We find that  the first two cumulants are sums of two contributions,
\begin{eqnarray}
\left\langle \nu_{a}\right\rangle  & = & \mu_{a}B+\sum_{c=a,b}\left(1+2\delta_{ac}\right)I_{c}F^{ac}p_{c},\label{eq:nu_1_two_species}\\
\left\langle \nu_{a}^{2}\right\rangle  & = & \sum_{c=a,b}\frac{1+8\delta_{ac}}{3}I_{c}\left(I_{c}+1\right)F_{2}^{ac}w\left(p_{c}\right),\label{eq:nu_2_two_species}
\end{eqnarray}
where $F^{aa}=g_{aa}\sum_{j}\frac{1}{r_{ij}^{3}}\left(3\gamma_{ij}^{2}-1\right)$
is a sum over sublattice $a$ and $F^{ab}=g_{ac}^{2}\sum_{j'}\frac{1}{r_{ij'}^{3}}\left(3\gamma_{ij'}^{2}-1\right)$
is over  sublattice $b$; both sums are evaluated in the
same reference frame. The polarisations of each species, $p_{a}$
and $p_{b}$, are kept as independent parameters.

The line shift Eq. (\ref{eq:nu_I_1}) can increase the line
width measured via the hyperfine shift of the electron Zeeman line
\cite{Gammon}. In this technique all nuclei are polarised first,
then they are depolarised at one frequency after another and 
shifts of the electron line are measured. But in the depolarisation process the
line position changes that can add to the measured line width. This addition
is anisotropic similarly to the zero polarisation line width in Eq. (\ref{eq:nu_I_2}). For Zinc Blende crystal
of a III-V semiconductor with $I_{a}=I_{b}=3/2$, using Eqs. (\ref{eq:nu_1_two_species}, \ref{eq:nu_2_two_species}) the maximum shift $\left(\left\langle \nu_{a}\right\rangle _{p=1}-\left\langle \nu_{a}\right\rangle _{p=0}\right)/\sqrt{\left\langle \nu_{a}^{2}\right\rangle }$
is $0$ when $B\parallel\left[001\right]$ and is $0.8$ when $B\parallel\left[111\right]$
for one sublattice \cite{ZincBlende} assuming that the other, off-resonance,
sublattice stays polarised.

\section{Single quantum of excitation above the  ferromagnetic state}
Now we  consider the situation where nuclei are close to fully polarised. The line width Eq. (\ref{eq:nu_I_2}) decreases at these large polarisations
because the exchange term in Eq. (\ref{eq:H0}) becomes dominant.
The eigenmodes of the exchange term, the magnons, are plane waves, so only a few, with long wavelengths, overlap significantly with the radio
frequency photon in Eq. (\ref{eq:mk}). In the limit of single
quantum of excitation above the fully ferromagnetic
state, corresponding to the subspace of $S_{z}=-IN+1$, the eigenmodes of full $H_{0}$
are plane waves.  As an example, for a cubic crystal they are $\left|\mathbf{q}\right\rangle =\sum_{j}\exp\left(\mathbf{q}\cdot\mathbf{r}_{ij}\right)I_{\mathbf{r}_{j}}^{+}\left|\Downarrow\right\rangle /\sqrt{2IN}$
where the momentum is quantised by the periodic boundary conditions
$\mathbf{q}=2\pi\mathbf{Q}/\left(L+1\right)^{d}$ and $0<Q_i<L+1$
are integers. The dispersion is obtained directly
using the Schroedinger equation, $H_{0}\left|\mathbf{q}\right\rangle =\left(E_{\mathbf{q}}+E_{1}\right)\left|\mathbf{q}\right\rangle $,
as $E_{\mathbf{q}}=-Ig\sum_{j:r_{j}>0}\cos\left(\mathbf{q}\cdot\mathbf{r}_{ij}\right)\left(3\gamma_{ij}^{2}-1\right)/r_{ij}^{3}$.
Here $E_{1}$ is a momentum independent part. The transition matrix
element in Eq. (\ref{eq:mk}) for the initial state $\left|f\right\rangle $
from the subspace $S_{z}=-IN$ is the largest for the smallest momentum
$q=2\pi/\left(L+1\right)^{d}$ and is suppressed for all other momenta
giving the zero line width in the limit $N \gg 1$.

In the subspace of $S^{z}=-IN+1$ the simple form of the eigenmodes
permits explicit calculation of the spin dynamics. The evolution of
a locally created polarisation, $\left|0\right\rangle =I_{\mathbf{L}/2}^{+}\left|\Downarrow\right\rangle /\sqrt{2I}$, due to the action of $H_{0}$, gives a spreading profile at later times, 
\begin{equation}
\left\langle t|I_{\mathbf{r}}^{z}|t\right\rangle =-I+\left|\sum_{\mathbf{q}}\frac{e^{i\mathbf{q}\cdot\left(\mathbf{r}-\frac{\mathbf{L}}{2}\right)-iE_{\mathbf{q}}t}}{N}\right|^{2}.\label{eq:Iz_evolution}
\end{equation}
For a large magnetic field, corrections to this average due to the
off-resonant part of the dipole interaction $H_{1}$ will involve
the two and the three excitation subspaces. In the specific case of
$d=1$ and only nearest neighbour interactions, the Bethe-Ansatz approach
\cite{GiamarchiBook} allows to assess these corrections, which turn out to be small, $\sim g/B$.

The width of the probability profile Eq. (\ref{eq:Iz_evolution})
grows in space linearly with time $r\sim tg/A^{2}$ \cite{AharonovDavidovichZagury},
where $g$ is  the dipolar coupling $g$ and $A$ is the lattice
parameter. At smaller polarisations $\left|p\right|<1$ the Ising
interaction leads to a magnon-magnon scattering which slows the free
magnon propagation. This sets a bound on the shortest nuclear polarisation
decay time for a quantum dot of a radius $r_{0}$ as $T_{1}=r_{0}A^{2}/g$.
The dipolar coupling constant can be extracted from the observed width
of the nuclear magnetic resonance. In a single crystal of GaAs the
full width at half maximum is $5$ kHz \cite{barrett}. Using Eq.
(\ref{eq:nu_2_two_species}) for zero polarisation and assuming $g_{\textrm{GaGa}}=g_{\textrm{AsAs}}=g_{\textrm{GaAs}}=g$
we extract $g=12\;\textrm{Hz}\cdot\textrm{nm}^{3}$ \cite{ZincBlende}.
For a lateral quantum dot of the radius $r_{0}=50$ nm the bound is thus
$T_{1}>1.25$ s. 

\section{Conclusions}
In conclusion, we have analysed the effect of a non-equilibirum polarisation on the nuclear magnetic resonance and have found a direct relation between the line shape  and  polarisation. This result suggests to use the line shape to measure directly the degree of polarisation achieved in dynamical nuclear pumping experiments. 

\section{Acknowledgement}
This work was supported by the EPSRC (UK) EP/G001642.
\appendix

\section{Third moment for $I=1/2$}

Here we evaluate the third cumulant as 

\begin{equation}
\left\langle \nu^{3}\right\rangle =\left\langle m^{3}\right\rangle -3\left\langle \nu^{2}\right\rangle \left\langle \nu\right\rangle -\left\langle \nu\right\rangle ^{3}.
\end{equation}
The first and the second cumulants are evaluated in
Eqs. (9,11). The third moment in Eq. (6)  involves
a product of two commutators 
\begin{equation}
\left[H_{0},I^{+}\right]  =  B\sum_{j}I_{j}^{+}+\sum_{i<j}\frac{g}{r_{ij}^{3}}3\left(3z_{ij}^{2}-1\right)I_{j}^{+}I_{i}^{z},\label{eq:commutator_1}
\end{equation}
\begin{widetext}
\begin{equation}
\left[H_{0},\left[H_{0},I^{-}\right]\right]  =  \sum_{i<j\neq j'}\frac{g}{r_{ij}^{3}}3\left(3\gamma_{ij}^{2}-1\right)\frac{g}{r_{jj'}^{3}}\left(3\gamma_{jj'}^{2}-1\right)\left(\frac{1}{2}I_{i}^{+}I_{j}^{-}I_{j'}^{-}-\frac{1}{2}I_{j}^{+}I_{i}^{-}I_{j'}^{-}+I_{j'}^{z}I_{j}^{z}I_{i}^{-}+2I_{j'}^{z}I_{i}^{z}I_{j}^{-}\right).\label{eq:commutator_2}
\end{equation}

We evaluate the trace of the operator parts of each term in the product
separately. Keep in mind that the terms with $i=j$ or $j=j'$ do
not enter into the sums thus we do not consider them. For the following
terms we add up contributions from all possible configurations of
three different spins

\begin{equation}
-\frac{1}{2}\textrm{Tr}\left(I_{j}^{+}I_{i}^{-}I_{j'}^{-}I_{j''}^{+}I_{i''}^{z}\right)=0,
\end{equation}
The above term is zero as we do not account for $j=i$ and $j=j'$
terms.
\begin{equation}
\frac{1}{2}\textrm{Tr}\left(I_{i}^{+}I_{j}^{-}I_{j'}^{-}I_{j''}^{+}I_{i''}^{z}\right)=\delta_{ij'}\delta_{jj''}\frac{1}{4}\left(C_{N-3}^{n-2}-C_{N-3}^{n-1}+\delta_{ii''}\left(C_{N-3}^{n-1}-C_{N-3}^{n-2}+C_{N-2}^{n-1}\right)\right),
\end{equation}
For the following terms we add up contributions from all possible
configurations of four different spins 
\begin{eqnarray}
\textrm{Tr}\left(I_{j'}^{z}I_{j}^{z}I_{i}^{-}I_{j''}^{+}I_{i''}^{z}\right) & = & \frac{\delta_{ij''}}{2^{3}}\bigg[\delta_{ji''}\big\{ C_{N-3}^{n-2}-C_{N-3}^{n}-C_{N-4}^{n-3}+C_{N-4}^{n}-3C_{N-4}^{n-1}+3C_{N-4}^{n-2}\big\}\\
 &  & +\delta_{j'i''}\big\{-C_{N-4}^{n-3}+C_{N-4}^{n}-3C_{N-4}^{n-1}+3C_{N-4}^{n-2}-C_{N-3}^{n}+C_{N-3}^{n-2}\big\}\nonumber \\
 &  & +\big\{ C_{N-4}^{n-3}-C_{N-4}^{n}+3C_{N-4}^{n-1}-3C_{N-4}^{n-2}\big\}\nonumber \\
 &  & +\delta_{ij'}\big\{ C_{N-4}^{n}-C_{N-3}^{n-2}-C_{N-4}^{n-3}-C_{N-3}^{n}-3C_{N-4}^{n-1}+2C_{N-3}^{n-1}+3C_{N-4}^{n-2}\big\}\nonumber \\
 &  & +\delta_{ji''}\delta_{ij'}\big\{-C_{N-2}^{n}-C_{N-2}^{n-1}-C_{N-4}^{n}+C_{N-4}^{n-3}+2C_{N-3}^{n}+3C_{N-4}^{n-1}-2C_{N-3}^{n-1}-3C_{N-4}^{n-2}\big\}\bigg],\nonumber 
\end{eqnarray}
\begin{eqnarray}
2\textrm{Tr}\left(I_{j'}^{z}I_{i}^{z}I_{j}^{-}I_{j''}^{+}I_{i''}^{z}\right) & = & \frac{\delta_{jj''}}{2^{2}}\bigg[\delta_{j'i}\delta_{ii''}\big\{ C_{N-4}^{n-3}-C_{N-4}^{n}+C_{N-2}^{n-1}-C_{N-2}^{n}-2C_{N-3}^{n-2}\\
 &  & +2C_{N-3}^{n}+3C_{N-4}^{n-1}-3C_{N-4}^{n-2}\big\}\nonumber \\
 &  & +\delta_{j'i}\big\{ C_{N-3}^{n-2}-C_{N-3}^{n}-C_{N-4}^{n-3}+C_{N-4}^{n}-3C_{N-4}^{n-1}+3C_{N-4}^{n-2}\big\}\nonumber \\
 &  & +\delta_{ii''}\big\{ C_{N-3}^{n-2}-C_{N-3}^{n}-C_{N-4}^{n-3}+C_{N-4}^{n}-3C_{N-4}^{n-1}+3C_{N-4}^{n-2}\big\}\nonumber \\
 &  & +\delta_{j'i''}\big\{ C_{N-3}^{n-2}-C_{N-3}^{n}-C_{N-4}^{n-3}+C_{N-4}^{n}-3C_{N-4}^{n-1}+3C_{N-4}^{n-2}\big\}\nonumber \\
 &  & +\delta_{j'i}\delta_{j'i''}\big\{-C_{N-3}^{n-2}+C_{N-3}^{n}+C_{N-4}^{n-3}-C_{N-4}^{n}+3C_{N-4}^{n-1}-3C_{N-4}^{n-2}\big\}\nonumber \\
 &  & +\delta_{j'i''}\delta_{ii''}\big\{ C_{N-4}^{n-3}-C_{N-4}^{n}+3C_{N-4}^{n-1}-3C_{N-4}^{n-2}\big\}\nonumber \\
 &  & +\big\{ C_{N-4}^{n-3}-C_{N-4}^{n}+3C_{N-4}^{n-1}-3C_{N-4}^{n-2}\big\}\nonumber \\
 &  & +\delta_{j'i''}\delta_{ii''}\delta_{j'i}\big\{-C_{N-4}^{n-3}+C_{N-4}^{n}-3C_{N-4}^{n-1}+3C_{N-4}^{n-2}\big\}\bigg].\nonumber 
\end{eqnarray}
Then, we substitute the combinatorial factors explicitly and divide
them by the denominator Eq. (7). Keeping only the
leading terms for $1/N\ll1$ we get four contributions to the third
moment,
\begin{eqnarray}
\frac{1}{2}\frac{\textrm{Tr}\left(I_{i}^{+}I_{j}^{-}I_{j'}^{-}I_{j''}^{+}I_{i''}^{z}\right)}{\textrm{Tr}\left(I^{-}I^{+}\right)} & = & \frac{\delta_{ij'}\delta_{jj''}}{4}\left[\frac{n\left(-N+2n\right)}{N^{3}}+\delta_{ii''}\frac{2n\left(N-n\right)}{N^{3}}\right],
\end{eqnarray}
\begin{eqnarray}
\frac{\textrm{Tr}\left(I_{j'}^{z}I_{j}^{z}I_{i}^{-}I_{j''}^{+}I_{i''}^{z}\right)}{\textrm{Tr}\left(I^{-}I^{+}\right)} & = & \frac{\delta_{ij''}}{8}\bigg[-\delta_{ji''}\frac{4n(2n^{2}-3nN+N^{2})}{N^{4}}-\delta_{j'i''}\frac{4n(2n^{2}-3nN+N^{2})}{N}\\
 &  & +\frac{\left(-N+2n\right)^{3}}{N^{4}}-\delta_{ij'}\frac{2n(-NN+2n)^{2}}{N^{4}}+\delta_{ji''}\delta_{ij'}\frac{8n^{2}(n-N)}{N^{4}}\bigg],\nonumber 
\end{eqnarray}
\begin{eqnarray}
2\frac{\textrm{Tr}\left(I_{j'}^{z}I_{i}^{z}I_{j}^{-}I_{j''}^{+}I_{i''}^{z}\right)}{\textrm{Tr}\left(I^{-}I^{+}\right)} & = & \frac{\delta_{jj''}}{4}\bigg[\delta_{j'i}\delta_{ii''}\frac{4n\left(2n^{2}-3nN+N^{2}\right)}{N^{4}}-\delta_{j'i}\frac{4n\left(2n^{2}-3nN+N^{2}\right)}{N^{4}}\\
 &  & -\delta_{ii''}\frac{4n\left(2n^{2}-3nN+N^{2}\right)}{N^{4}}-\delta_{j'i''}\frac{4n\left(2n^{2}-3nN+N^{2}\right)}{N^{4}}\nonumber \\
 &  & +\delta_{j'i}\delta_{j'i''}\frac{4n\left(2n^{2}-3nN+N^{2}\right)}{N^{4}}+\delta_{j'i''}\delta_{ii''}\frac{\left(-N+2n\right)^{3}}{N^{4}}\nonumber \\
 &  & +\frac{\left(-N+2n\right)^{3}}{N^{4}}-\delta_{j'i''}\delta_{ii''}\delta_{j'i}\frac{\left(-N+2n\right)^{3}}{N^{4}}\bigg].\nonumber 
\end{eqnarray}
\end{widetext}
Substituting these expressions in to the sums Eqs. (\ref{eq:commutator_1},\ref{eq:commutator_2})
and identifying $p=-1+2n/N$ we obtain Eq.
(12).

\section{Derivation of $_{I}A_{n}^{N}$ for arbitrary $I$}

Here we derive the expression for the combinatorial coefficient for
arbitrary $I$ which uses only one sum over the binomial factors.
We start from the generating function for a single spin 
\begin{equation}
f\left(x\right)=1+x+\dots+x^{2I}\equiv\frac{1-x^{2I+1}}{1-x}.
\end{equation}
Then the combinatorial factor is given by the coefficient in front
of the $n^{th}$ power of $x$ in 
\begin{eqnarray}
f^{N}\left(x\right) & = & \sum_{k_{1}=0}^{N}C_{k_{1}}^{N}\left(-1\right)^{k_{1}}x^{\left(2I+1\right)k_{1}}\sum_{k_{2}=0}^{\infty}C_{k_{2}}^{N-1+k_{2}}x^{k_{2}}\nonumber \\
 & = & \sum_{k_{1},k_{2}=0}^{\infty}\left(-1\right)^{k_{1}}C_{k_{1}}^{N}C_{k_{2}}^{N-1+k_{2}}x^{\left(2I+1\right)k_{1}+k_{2}},
\end{eqnarray}
where $\left(1+x\right)^{N}=\sum_{k=0}^{N}C_{k}^{N}x^{k}$ and $\left(1-x\right)^{-N}=\sum_{k=0}^{\infty}C_{k}^{N-1+k}x^{k}$
were used. Here the coefficient is 
\begin{equation}
_{I}A_{n}^{N}=\sum_{k=0}^{n/\left(2I+1\right)}\left(-1\right)^{k}C_{k}^{N}C_{n-\left(2I+1\right)k}^{N-1+n-\left(2I+1\right)k}.
\end{equation}

\section{Numerical evaluation of the first two moments for $I>1/2$}

\begin{figure}
\centering\includegraphics[width=0.95\columnwidth]{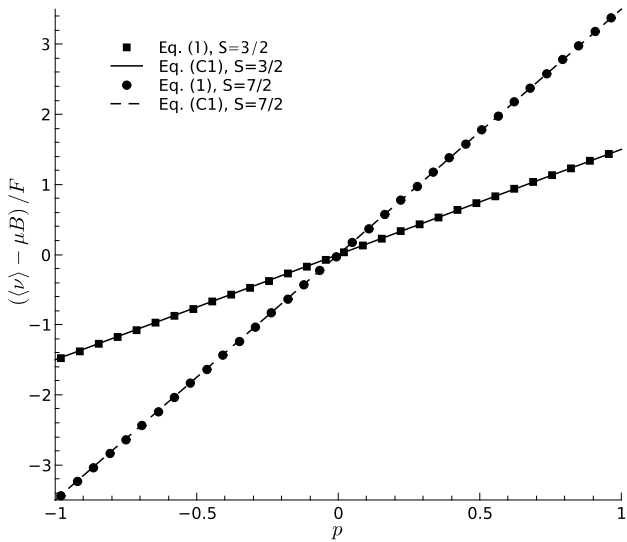}

\caption{The first moment of the line shape. The full and dashed lines are
linear function from Eq. (1)  for $I=3/2$ and $I=5/2$.
The squares and ellipses are Eq. (\ref{eq:nu_I}) numerically evaluated
using for $I=3/2$ and $I=5/2$; $N=500$.\label{fig:nu}}
\end{figure}

\begin{figure}
\centering\includegraphics[width=0.95\columnwidth]{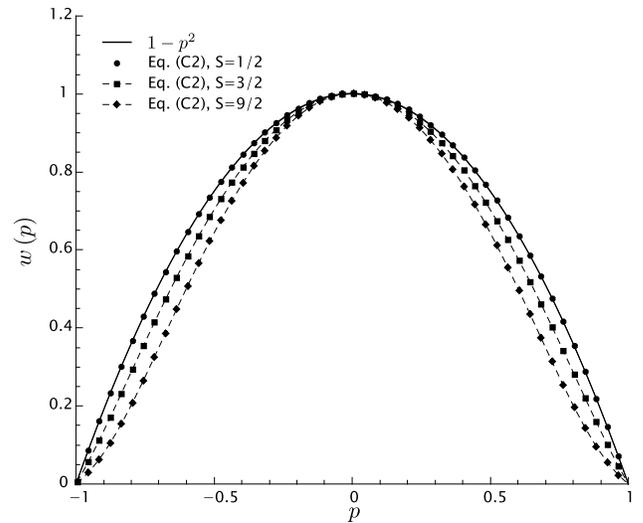}
\caption{The second moment of the line shape. The full line is $w\left(p\right)=1-p^{2}$.
The ellipses, squares, and diamonds are Eq. (\ref{eq:nu2_I}) numerically
evaluated for $I=1/2$, $I=3/2$ and $I=9/2$; $N=500$.\label{fig:nu2}}
\end{figure}

For a nuclear spin $I>1/2$ the cumulants of the line shape are expressed
in terms of sums over the binomial coefficients. We analyse these
sums numerically. Substituting Eqs. (13,14)  into
Eq. (6), we obtain the first moment,

\begin{widetext}
\begin{equation}
\frac{\left\langle \nu\right\rangle -\mu B}{F}=\frac{\sum_{k_{1},k_{2}=0}^{2I}{}_{I}A_{n-k_{1}-k_{2}}^{N-2}\left(-I+k_{1}\right)\left(k_{2}+1\right)\left(2I-k_{2}\right)}{\sum_{k=0}^{2I}{}_{I}A_{n-k}^{N-1}\left(k+1\right)\left(2I-k\right)}.\label{eq:nu_I}
\end{equation}
This expression coincides with Eq. (1)  when $N\gg1$,
see Figure \ref{fig:nu}. 

The polarisation dependence, $w\left(p\right)=\left\langle \nu^{2}\right\rangle /\left(3I\left(I+1\right)F_{2}\right)$,
of the second cumulant is obtained by substituting Eqs. (12,13) into Eq. (6),
\begin{eqnarray}
w\left(p\right) & = & -\frac{3\sum_{k_{1},k_{2},k_{3}=0}^{2I}{}_{I}A_{n-k_{1}-k_{2}-k_{3}}^{N-3}\left(k_{1}+1\right)\left(2I-k_{1}\right)\left(-I+k_{2}\right)\left(-I+k_{3}\right)}{\sum_{k=0}^{2I}{}_{I}A_{n-k}^{N-1}\left(k+1\right)\left(2I-k\right)}\nonumber \\
 &  & +\frac{3\sum_{k_{1},k_{2},k_{3}=0}^{2I}{}_{I}A_{n-k_{1}-k_{2}-k_{3}}^{N-3}\left(k_{1}+1\right)\left(2I-k_{1}\right)\left(-I+k_{2}\right)\left(-I+k_{3}\right)}{\sum_{k=0}^{2I}{}_{I}A_{n-k}^{N-1}\left(k+1\right)\left(2I-k\right)}.\label{eq:nu2_I}
\end{eqnarray}
\end{widetext}
In the limit $1/N\ll1$ this expression coincides with $w\left(p\right)=1-p^{2}$
for $I=1/2$. When $I>1/2$ the narrowing is slightly faster than
$1-p^{2}$, see comparison in Figure. \ref{fig:nu2}. 


\begin{thebibliography}{10}
\bibitem{ZincBlende}For Zinc Blende crystal the parameters in Eq.
(\ref{eq:nu_2_two_species}) are $F^{aa}=0$, $F_{2}^{aa}=68.2g_{aa}^{2}/A^{6}$,
$F_{2}^{ab}=49.6g_{ab}^{2}/A^{6}$ for
$B\parallel\left[001\right]$ and $F^{aa}=6.2g_{aa}/A^{3}$, $F_{2}^{aa}=105.1g_{aa}^{2}/A^{6}$
for $B\parallel\left[111\right]$. The lattice constant of GaAs is
$A=0.56$ nm.
\bibitem{skewness_cubic}For cubic crystal and $\mathbf{B}\parallel\left[001\right]$
the parameters in Eq. (\ref{eq:nu_3_12}) are $F_{3c}-F_{2}F-5F_{3}=-83.2g^{3}/A^{9}$
and $F_{2}=13.4g^{2}/A^{6}$ where $A$ is the lattice constant.
\end{thebibliography}
\end{document}